\documentclass[11pt]{article}
\usepackage{amsmath}
\usepackage{amssymb}
\usepackage{amsfonts}
\usepackage{eucal}
\usepackage{latexsym}
\usepackage{a4wide}
\usepackage{natbib}
\usepackage{eucal}
\begin{document}

%\draft

%%%%%%%%%%%%%%%%%%%%%%%%%%%%%%%%%%

\newcommand{\pot}[2]{#1\times10^{#2}}

\newcommand{\beal}{\begin{align}}
\newcommand{\be}{\begin{eqnarray}}
\newcommand{\ee}{\end{eqnarray}}
\newcommand{\bse}{\begin{subequations}}
\newcommand{\ese}{\end{subequations}}

\newcommand{\bs}{\boldsymbol}
\newcommand{\mbb}{\mathbb}
\newcommand{\mcal}{\mathcal}
\newcommand{\mfr}{\mathfrak}
\newcommand{\mrm}{\mathrm}

\newcommand{\ovl}{\overline}
\newcommand{\p}{\partial}
\newcommand{\f}{\frac}
\newcommand{\fa}{\forall\;}
\newcommand{\diff}{\mrm{d}}
\newcommand{\lan}{\bigl\langle}
\newcommand{\ran}{\bigl\rangle}
\newcommand{\vol}{\mrm{vol}}

\newcommand{\R}{\mathbb{R}}
\newcommand{\N}{\mathbb{N}}
\newcommand{\Z}{\mathbb{Z}}

\newcommand{\Ham}{\hat{H}}
\newcommand{\HH}{\hat{\mathcal{H}}}
\newcommand{\T}{\bs{\mathcal{T}}}

\newcommand{\rt}{\tilde{r}}
\newcommand{\st}{\tilde{s}}

\newcommand{\ga}{\alpha}
\newcommand{\gb}{\beta}
\newcommand{\gc}{\gamma}
\newcommand{\Gc}{\Gamma}
\newcommand{\gd}{\delta}
\newcommand{\Gd}{\Delta}
\newcommand{\gl}{\lambda}
\newcommand{\gf}{\varphi}
\newcommand{\Gf}{\Phi}
\newcommand{\gk}{\kappa}
\newcommand{\vgk}{\varkappa}
\newcommand{\go}{\omega}
\newcommand{\Go}{\Omega}
\newcommand{\veps}{\varepsilon}
\newcommand{\eps}{\epsilon}
\newcommand{\vphi}{\varphi}
\newcommand{\gr}{\varrho}
\newcommand{\gs}{\sigma}
\newcommand{\Gs}{\Sigma}
\newcommand{\gt}{\theta}
\newcommand{\vgt}{\vartheta}

\newcommand{\csp}{\;,\qquad}

%%%%%%%%%%%%%%%%%%%%%%%%%%%%%%%%%%%%%%%%%%%%%%%%%%%%%%%%%%%%%%%%%%%%%%
\title{Accretion of helium and metal-rich gas onto neutron stars and
black holes at high luminosities}

\author{J\"orn Dunkel$^1$\footnote{\texttt{dunkel@mpa-garching.mpg.de}}, Jens Chluba$^1$, and Rashid A. Sunyaev$^{1,2}$}
\date{\today}

\maketitle
$^1$Max-Planck-Institute for Astrophysics,
Karl-Schwarzschild-Stra{\ss}e 1, 85741 Garching, Germany\\
$^2$ Space Research Institute, Russian Academy of Sciences,
Profsoyuznaya 84/32, 117997 Moscow, Russia

\begin{abstract}
Ultraluminous X-ray sources fed by Wolf-Rayet star winds and
X-ray bursters in ultracompact binaries with
He or C white dwarfs have accretion disks, whose properties may significantly
differ from those of pure H $\ga$-accretion disks. We have therefore included the dependence
on charge number $Z$ and mean molecular weights $\mu_{e/I}$ into the \citet{ShSu73} scaling relations for the key parameters of
the disk. Furthermore, we also consider the case of the pseudo-Newtonian
potential of \cite{PaWi80}. These scaling
relations might become useful, e.g., when estimating the illumination
efficiency of the external parts of the disk. We also address the changes in the structure of the boundary (spreading) layer on the
surface of neutron stars, occurring in the case of H depleted accretion disks.
\end{abstract}
          %97.60.Gb Pulsars) 
98.62.Mw  %Infall, accretion, and accretion disks (see also 04.70.-s
          %Physics of black holes in general relativity and
          %gravitation) 

%%%%%%%%%%%%%%%%%%%%%%%%%%%%%%%%%%%%%%%%%%%%%%%%%%%%%%%%%%%%%%%%%%%%%%

%%%%%%%%%%%%%%%%%%%%%%%%%%%%%%%%%%%%%%%%%%%%%%%%%%%%%%%%%%%%%%%%%%%%%%
\section{Introduction}
\label{s:intro}
%%%%%%%%%%%%%%%%%%%%%%%%%%%%%%%%%%%%%%%%%%%%%%%%%%%%%%%%%%%%%%%%%%%%%%

Recent observations of galaxies by the CHANDRA
X-ray observatory confirm the
existence of compact, ultraluminous X-ray sources (ULXs), emitting
at luminosities considerably higher than the Eddington luminosities of
neutrons stars or stellar mass black holes with $1-15\; M_\odot$ \citep{RoEtAl04a,RoEtAl04b,LiMi05}.  The most dramatic results
were obtained by CHANDRA observations of star forming galaxies and
regions \citep{GrGiSu03}, where over the large luminosity range $L=10^{36}-10^{40}$ erg/sec the corresponding high mass X-ray binary
(HMXB) luminosity function is well fitted by a single power law with  
a strong cut-off at $L\sim 10^{40.5}$ erg/sec, which is
more than  two orders of magnitude higher than the Eddington
luminosity of a solar mass neutron star. Due to the fact that the HMXB
luminosity function does not exhibit a change of slope or other
peculiarities in the region just above the Eddington limit of stellar
mass black holes, it is plausible to assume that ULXs represent the
high-mass/high-accretion-rate tail of the \lq ordinary\rq\space HMXB
population with masses in the range  $10-15\; M_\odot$ (King, 2002;
Grimm et al. 2002; Gilfanov, 2004). 
\par
These new observational results increased strongly the
theoretical interest in the question, whether an accreting object can
produce super-Eddington luminosities, or whether the Eddington
luminosity provides a real upper limit for the luminosity of accreting
objects. In principle,
one can think of several different mechanisms leading to
super-Eddington luminosities and, at present,  there does not yet exist a
commonly accepted model for the huge energy output of ULXs. For example, young
rotation powered pulsars or jet sources directed towards us could
substantially increase the observed X-ray flux. Other possibilities 
might be slim accretion disks \citep{AbCzLaSz88}, or an accretion disk
that experiences radiation driven
inhomogeneities \citep{Be02}. Alternatively, many observers and
theorists believe that the high luminosity of ULXs is due to accretion onto an
intermediate mass black hole with $10^2-10^4\;M_\odot$ \citep{MiCo04,KuDoMa02}. 
\par
On the other hand, there also exist several \lq minor\rq\space effects (e.g.,
related to a disk's inclination and chemical composition), each of
which being, in principle, able to increase the  observable luminosity
of an accreting object by a factor of two or three \citep{GrGiSu02}. Due to the
aforementioned homogeneity of the HMXB luminosity function, it is
worthwhile to carefully reconsider the contributions of these \lq
minor\rq\space effects, even though most of them are known to
theorists for more than 30 years. In the present paper we are going to focus on the case, where the chemical composition of the accreting matter  
is strongly deviating from the standard cosmic
abundances. Obviously, in this case the mass per electron in the
accreting gas can be higher than for a pure H plasma or
standard cosmic abundances.  It is
well-known that for completely ionized He, C, O, N, or Mg plasmas,
having twice less electrons per baryon than H, the Eddington luminosity
is by a factor of two higher than for a pure H plasma. 
\par
From the observational point of view there are evidences that,
under certain conditions, accreting gas in binary systems can consist
only of elements heavier than H \citep{HaEtAl05}. For example, several observations of 
{\em X-ray sources in extremely dense binary systems} revealed a neutron
star accreting matter supplied by a He, C, O, Ne, or even Mg
white dwarf \citep{ScEtAl01,JuCh03,NeEtAl04}. The unusual chemical abundance influences
the power release due to nuclear explosions on the surface of the
neutron star, which are observable as X-ray bursts. The duration of
bursts and their repetition rate are also affected strongly. 
\par
The problem of non-standard abundances might also be
of relevance, if the accretion onto the black hole of the HMXB takes place
in a {\em star forming region or galaxy}. Then many of the HMXBs
are fed by stellar wind accretion. A particular interesting scenario in this
regard corresponds to the case, where the donor star is of
Wolf-Rayet-type. During their evolution process these stars may have lost a
significant amount of their H rich envelope, leading to stellar
winds dominated by He or, in some case, even by C and N
\citep{HaKo01,Hu01,CrEtAl02,AbEtAl04}. For instance, one of 
the brightest, most luminous X-ray sources in the Milky Way, Cygnus X-3, is
fed by the dense, high-velocity wind from its He Wolf-Rayet companion  
(van Kerkwijk et al., 1992; Lommen et al., 2005). 
\par
In the case of accretion onto a black hole, the plasma density in the
region of the main energy release of a standard H dominated accretion
disk is not high enough to produce the photons necessary for creating a
black body radiation spectrum inside the disk \citep{ShSu73}. Moreover,
in the limit  $\dot M/\dot M_\mrm{E} \to 1$ and $\ga\to 1$ the optical
depth becomes rather small (here, as usual, $\dot M_\mrm{E}$ denotes
the accretion rate corresponding to the Eddington luminosity and $\ga$
is the viscosity parameter). It is, therefore, the purpose of the
present paper to study how the disk's density, temperature, and
optical depth depend on the chemical abundance  in the extreme case of
 H poor  accretion.  By simply repeating the approach of
Shakura and Sunyaev (1973), we will show that under such peculiar
conditions the density of electrons and nuclei increases, thereby also
increasing the effective optical depth. This might 
stabilize the disk against the transition to the two-temperature hot flow
regime \citep{ShLiEa76}. 
\par
Recently, \citet{HaEtAl05} have performed 
numerical simulations to determine the structure in the outer regions of
C/O/Ne-dominated accretion disks in ultracompact binary systems (for
details of their NLTE-model see Nagel et al., 2004). Their primary
objective was to explain the 
observed peculiarities in the optical spectra of these systems, by 
taking into account metal-line blanketing and irradiation by the
central object. However, in spite of considerable theoretical
interest in the properties of H poor accretion disk \citep{MePeHe02}, we were unable to find in the literature
simple analytical results, describing the power law scaling relations
for key parameters of the standard accretion disk with respect to its
chemical composition in the extreme case of H or even He depleted accretion flows. Therefore,
the formulae given below, which have originally been calculated as
part of a more complex study, will possibly be useful when
interpreting observational data and deriving simple analytical estimates. Finally, it is also 
worthwhile to mention that a high abundance of He or heavier elements
does not only decrease the disk (scale) height by a factor of about
two, but also changes the width and the surface brightness of the
spreading layer on the surface of neutron stars. Therefore, in the periods between X-ray bursts, the
accretion of heavier elements also influences the heating 
of the outer parts of the disk by X-ray radiation emitted from
the central disk regions (this also applies to the case of accretion
onto black holes) and
from the surface of the neutron star \citep{LySu76}.

%%%%%%%%%%%%%%%%%%%%%%%%%%%%%%%%%%%%%%%%%%%%%%%%%%%%%%%%%%%%%%%%%%%%%%
\section{Structure of helium or metal dominated accretion disks}

The hydrodynamic equations considered by \citet{ShSu73} do not explicitly depend on the
chemical composition of the disk. More precisely, in their model the chemical
composition enters only via the thermodynamic equations for the pressure
and the energy density of the gas:
\bse\label{e:mod}
\be\label{e:eos}
P_g=\f{\gr k T}{\mu m_p},\qquad
\eps_g=\f{3}{2}\,\f{\gr k T}{\mu m_p},
\ee
and via the Rosseland mean opacities for Thomson and free-free processes:
\be
\gk_\mrm{T}&=&\f{\gs_\mrm{T}}{\mu_e m_p}
=\f{0.398}{\mu_e}\; \mrm{cm^2\;g^{-1}},\\
\gk_\mrm{ff}&=&0.12 \;
\f{Z^2}{\mu_e \mu_I} 
\left(\f{\gr\;\mrm{cm^3}}{m_p}\right)
\left(\f{T}{\mrm{K}}\right)^{-7/2}
\mrm{\;cm^2\;g^{-1}}.\quad\label{e:free}
\ee
\ese
Here $Z$ denotes the charge number of the ions, $\rho$ the mass
density of matter, $k$ the Boltzmann constant, $\gs_\mrm{T}$ the
Thomson scattering cross section, $T$ the temperature,
$m_p$ the proton mass, $\mu_{e/I}$ the mean molecular weight of the
electrons and ions, respectively, and $\mu^{-1}=
\mu_e^{-1}+\mu_I^{-1}$. The results for the disk's scale height,
surface density, etc., given below in 
Eqs. \eqref{e:zone_A} -- \eqref{e:zone_C}, were obtained by using
Eqs. \eqref{e:mod} and following the approach of \citet{ShSu73}. In particular, Eqs. \eqref{e:zone_A} -- \eqref{e:zone_C} refer
to a Newtonian potential $\Phi=-{G M}/(R^2+z^2)^{1/2}$, where $R$ is the
cylindrical radius. It is however easy to see that the scaling
relations with respect to the chemical parameters  (see also Table \ref{table_1}) remain also valid, when
including relativistic corrections in the form of the pseudo-Newtonian
potential of \citet{PaWi80}, as evident from the results in Appendix A. It is convenient to
express the results in terms of non-dimensional quantities
$m={M}/{M_\odot},\;\dot{m}=\dot{M}/{\dot M_\mrm{E}},\;r=R/r_0$ and $s=1-r^{-1/2},$ 
where $r_0=3r_G=6GM/c^2= m\; 8.86\; \mrm{km}$.  The total luminosity of an
infinite $\ga$-disk, ranging from $r_0$ to $\infty$ reads $L_\infty= \zeta
\dot{M} c^2$ with an efficiency $\zeta=1/12$ in the case of a Newtonian
potential and $\zeta\approx 0.06$ for a Schwarzschild black hole.
The critical Eddington luminosity, providing an upper limit for
spherically symmetric accretion, can then be written as 
\be\label{e:Eddington_lum}
L_\mrm{E}
=\mu_e L_\mrm{E,H}
=\mu_e\,\f{2\pi r_G\,m_pc^3}{\gs_\mrm{T}}
=\mu_e\, m\;(\pot{1.26}{38}\,\mrm{erg\; sec}^{-1}),
\ee
where, here and below, quantities labelled by H refer to the case of  a fully
ionized, pure H plasma. 
The corresponding critical accretion rate, obtained from the
condition $L_\infty=L_\mrm{E}$, reads
\be
\label{e:Eddington_rate}
\dot M_\mrm{E}= \mu_e\dot M_\mrm{E,H}
=\mu_e\,\zeta^{-1}\,m\; (\pot{2.22}{-9}\,M_\odot\; \mrm{year}^{-1}).
\ee
%%%%%%%%%%%%%%%%%%%%%%%%%%%%%%%%%%%%%%%%%%%%%%%%%%
\begin{table}[p]
\centering
\begin{tabular}{llrrrrrr}
\hline
\hline
&I$^{Z+}$ & He$^{2+}$& C$^{6+}$& N$^{7+}$& 0$^{8+}$& Mg$^{12+}$\\
\hline
 &$Z$      &  2 &6     & 7    & 8       & 12      \\
 &$\mu$    & 4/3      & 12/7  & 14/8 & 16/9    & 24/13   \\
 &$\mu_e$  & 2        & 2     & 2    & 2       & 2       \\
 &$\mu_I$  & 4        & 12    & 14   & 16      & 24      \\
 &$\dot M_\mrm{E} /\dot M_\mrm{E,H}$ 
         & 2       & 2       & 2    & 2       & 2        \\
\hline
zone A: 
 &$H/H_\mrm{H}$
         & 0.5      & 0.5   &  0.5 & 0.5     & 0.5       \\
 &$\Gs/\Gs_\mrm{H}$
         & 4        & 4     &  4   & 4       & 4        \\
 &$\gr_c/\gr_{c,\mrm{H}}$
         & 8        & 8     &  8   & 8       & 8        \\
 &$n_{ec}/n_{ec,\mrm{H}}$
         & 4        & 4     &  4   & 4       & 4        \\
 &$\bar v_R/\bar v_{R,\mrm{H}}$
         & 0.25     & 0.25  &  0.25 & 0.25   & 0.25        \\
 &$\eps_{\gc c}/\eps_{\gc c,\mrm{H}}$
         & 2        & 2     &  2   & 2       & 2        \\
 &$P_c/P_{c,\mrm{H}}$
         & 2        & 2     &  2   & 2       & 2        \\
 & $T_c/T_{c,\mrm{H}}$
         & 1.19     & 1.19  & 1.19 & 1.19    & 1.19   \\
 & $\tau_c/\tau_{c,\mrm{H}}$
         & 4.18     & 7.23  & 7.81 & 8.35    & 10.23  \\
 & $t_\mrm{co}/t_\mrm{co,H}$
         & 0.5  & 0.5   & 0.5   & 0.5   & 0.5    \\
\hline
zone B: 
 &$H/H_\mrm{H}$
         & 0.63     & 0.57  & 0.57  & 0.56    & 0.55    \\
 &$\Gs/\Gs_\mrm{H}$
         & 2.51     & 3.08  & 3.13  & 3.17    & 3.27     \\
 &$\gr_c/\gr_\mrm{c,H}$
         & 3.99     & 5.40  & 5.54  & 5.64    & 5.90     \\
 &$n_{ec}/n_{ec,\mrm{H}}$
         & 2.00     & 2.70  & 2.77  & 2.82    & 2.95     \\
 &$\bar v_R/\bar v_{R,\mrm{H}}$
         & 0.40     & 0.32  & 0.32  & 0.32    & 0.31     \\
 &$\eps_{\gc c}/\eps_{\gc c,\mrm{H}}$
         & 1.26     & 1.54  & 1.56  & 1.58    & 1.63     \\
 &$P_c/P_{c,\mrm{H}}$
         & 1.59     & 1.75  & 1.77  & 1.78    & 1.81     \\
 &$T_c/T_{c,\mrm{H}}$
         & 1.06     & 1.11  & 1.12  & 1.12    & 1.13    \\
 & $\tau_c/\tau_{c,\mrm{H}}$
         & 2.27     & 5.13  & 5.66  & 6.15    & 7.84   \\
 & $t_\mrm{co}/t_\mrm{co,H}$
         & 0.79     & 0.65  & 0.64  & 0.63    & 0.61     \\
\hline
zone C: 
 &$H/H_\mrm{H}$
         & 0.64     & 0.64  & 0.64 & 0.64    & 0.65     \\
 &$\Gs/\Gs_\mrm{H}$
         & 2.24     & 2.42  & 2.42 & 2.42    & 2.39     \\
 &$\gr_c/\gr_{c,\mrm{H}}$
         & 3.34     & 3.76  & 3.76 & 3.76    & 3.69     \\
 &$n_{ec}/n_{ec,\mrm{H}}$
         & 1.67     & 1.88  & 1.88 & 1.88    & 1.84     \\
 &$\bar v_R/\bar v_{R,\mrm{H}}$
         & 0.45     & 0.41  & 0.41 & 0.41    & 0.42     \\
 &$\eps_{\gc c}/\eps_{\gc c,\mrm{H}}$
         & 2.02     & 4.03  & 4.38 & 4.69    & 5.73    \\
 &$P_c/P_{c,\mrm{H}}$
         & 1.50     & 1.56  & 1.56 & 1.55    & 1.54     \\
 & $T_c/T_{c,\mrm{H}}$
         & 1.19     & 1.41  & 1.45 & 1.47    & 1.55     \\
 & $\tau_c/\tau_{c,\mrm{H}}$
         & 1.50     & 2.21  & 2.30 & 2.38    & 2.61     \\
 & $t_\mrm{co}/t_\mrm{co,H}$
         & 0.50     & 0.25  & 0.23 & 0.21    & 0.17      \\
\hline
\end{tabular}
\caption{Relative quantitative changes compared to a pure, fully
ionized H accretion disk with $\mu_I=\mu_e=Z=1$ and same parameter values
$(\ga,M,\dot{M})$.  Inner zone A:
$P_g\ll P_\gc$ and $\gk_\mrm{T} \gg \gk_\mrm{ff}$.  Intermediate zone B: $P_g\gg P_\gc$ and $\gk_\mrm{T}
\gg \gk_\mrm{ff}$.  Outer zone C: $P_g\gg P_\gc$ and $\gk_\mrm{T} \ll
\gk_\mrm{ff}$.  All values refer to pure disk plasmas, containing only
a single ion species.  The same results are obtained for the
pseudo-Newtonian potential of \citet{PaWi80}, see
Appendix A.
\label{table_1}} 
\end{table}
%%%%%%%%%%%%%%%%%%%%%%%%%%%%%%%%%%%%%%%%%%%%%%%%%%
For the inner disk zone (A),
characterized by dominating radiation pressure $P_g\ll P_\gc$ and $\gk_\mrm{T} \gg\gk_\mrm{ff}$, one finds 
%{\small
\be\label{e:zone_A}
H &=& 
\zeta^{-1}\;  m \;\dot m\;s
\;(2.2\;\mrm{km})\\
\Gs &=&\notag
\zeta\;\mu_e\;\ga^{-1}\;
\dot m^{-1}\;r^{3/2}\;s^{-1}\;
(9.9\times 10^2\;\mrm{g\;cm^{-2}})\\
\gr_c &=&\notag
\zeta^2\;\mu_e\;
\ga^{-1}\; m^{-1}\;
\dot{m}^{-2}\; r^{3/2}\;s^{-2}\;
%\;\\&&\notag
(2.2\times 10^{-4}\;\mrm{g\;cm^{-3}})\\
{n_{ec}} &=&\notag
\zeta^2\;\ga^{-1}\; m^{-1}\;\dot{m}^{-2}\; r^{3/2}\;s^{-2}\;
(1.3\times 10^{20}\;\mrm{cm^{-3}})\\
\bar v_R &=&\notag
\zeta^{-2}\;\ga\;\dot m^2\;r^{-5/2}\;s\;
(-2.5\times 10^3\;\mrm{km\; sec^{-1}})\\
\eps_{\gc c} &=&\notag
\mu_e\;\ga^{-1}\;m^{-1}\;r^{-3/2}\;
(3.1\times 10^{15}\;\mrm{erg\;cm^{-3}})\\
P_c &=&\notag
\mu_e\;\ga^{-1}\;m^{-1}\;r^{-3/2}\;
(1.0\times 10^{15}\;\mrm{erg\;cm^{-3}})\\
T_c &=&\notag
\mu_e^{1/4}\;\ga^{-1/4}\;m^{-1/4}\;r^{-3/8}\;
(2.5\times 10^{7}\;\mrm{K})\\
\tau_c &=&\notag
Z\;\zeta^2\; \mu_e^{1/16}\;\mu_I^{-1/2}\;\ga^{-17/16}\\
&&\notag
m^{-1/16}\;\dot{m}^{-2}\;r^{93/32}\; s^{-2}
\;(1.4\times 10^{-2}),
\ee
%}
where we note that $\dot m=\dot m_\mrm{H}/\mu_e$. 
Here the index $c$ is used to label midplane values, and $H$ is the
disk scale height, $\Gs$ the surface mass density, $\bar{v}_R$ the
(vertically averaged) radial velocity, $\eps_{\gc c}$ the energy
density of radiation,  $n_{ec}= \gr_c/(\mu_e m_p)$ the number density
of electrons, and
$\tau=\left(\gk_\mrm{T}\gk_\mrm{ff}\right)^{1/2}\Gs/2$ the effective
optical depth \citep{ShSu73}. 
\par
Similarly, for the intermediate zone (B) with $P_g\gg P_\gc$
and $\gk_\mrm{T} \gg \gk_\mrm{ff}$:
%{\small
\be\label{e:zone_B}
H&=& 
\zeta^{-1/5}\;\mu^{-2/5}\;\mu_e^{1/10}\;\ga^{-1/10}\;\\
&& \notag
m^{9/10}\;\dot m^{1/5}\;r^{21/20}\;s^{1/5}
\;(1.0\times 10^{-1} \;\mrm{km})\\
\Gs&=&\notag
\zeta^{-3/5}\;\mu^{4/5}\;\mu_e^{4/5}\;\ga^{-4/5}\;\\
&&\notag m^{1/5}\;\dot m^{3/5}\;r^{-3/5}\;s^{3/5}\;
(4.7\times 10^4\;\mrm{g\;cm^{-2}})\\
\gr_c&=&\notag
\zeta^{-2/5}\;\mu^{6/5}\;\mu_e^{7/10}\;\ga^{-7/10}\;\\
&&\notag 
m^{-7/10}\;\dot{m}^{2/5}\; r^{-33/20}\;s^{2/5}
\;(2.3\;\mrm{g\;cm^{-3}})\\ 
n_{ec}&=&\notag
\zeta^{-2/5}\;\mu^{6/5}\;\mu_e^{-3/10}\;\ga^{-7/10}\\
&&\notag m^{-7/10}\;\dot{m}^{2/5}\; r^{-33/20}\;s^{2/5}
\;(1.4\times 10^{24}\; \mrm{cm^{-3}})\\ 
\bar v_R&=&\notag
\zeta^{-2/5}\;\mu^{-4/5}\;\mu_e^{1/5}\;\ga^{4/5}\; \\
&&\notag m^{-1/5}\;\dot{m}^{2/5}\; r^{-2/5}\;s^{-3/5}
\;(- 5.3\; \mrm{km\; sec^{-1}})\\ 
\eps_{\gc c}&=&\notag
\zeta^{-8/5}\;\mu^{4/5}\;\mu_e^{4/5}\;\ga^{-4/5}\; \\
&&\notag m^{-4/5}\;\dot m^{8/5}\;r^{-18/5}\;s^{8/5}
\;(1.5\times 10^{18}\;\mrm{erg\;cm^{-3}}) \\
P_c&=&\notag
\zeta^{-4/5}\;\mu^{2/5}\;\mu_e^{9/10}\;\ga^{-9/10}\;\\ 
&&\notag m^{-9/10}\;\dot m^{4/5}\; r^{-51/20}\;s^{4/5}
\;(2.3\times 10^{16}\;\mrm{erg\;cm^{-3}}) \\
T_c&=&\notag
\zeta^{-2/5}\;\mu^{1/5}\;\mu_e^{1/5}\;\ga^{-1/5}\; \\
&&\notag m^{-1/5}\;\dot m^{2/5}\; r^{-9/10}\;s^{2/5}
\;(1.2\times 10^8\;\mrm{K})\\
\tau_c&=&\notag
Z\;\zeta^{-1/10}\; \mu^{21/10}\;\mu_e^{-1/5}\;\mu_I^{-1/2}\;
\ga^{-4/5}\;\\
&&\notag
m^{1/5}\;\dot{m}^{1/10}\;r^{3/20}\; s^{1/10}\;
(4.5\times 10^1),
\ee
%}
and for the outer zone (C) with $P_g\gg P_\gc$ and $\gk_\mrm{T} \ll 
\gk_\mrm{ff}$: 
%{\small
\be\label{e:zone_C}
 H&=& 
Z^{1/10}\;\zeta^{-3/20}\;\mu^{-3/8}\;\mu_e^{1/10}\;\mu_I^{-1/20}\;
\ga^{-1/10}\; \\&&\notag
m^{9/10}\;\dot m^{3/20}\;r^{9/8}\;s^{3/20}\;
(6.1 \times 10^{-2}\;\mrm{km}) \\
\Gs&=&\notag
Z^{-1/5}\;
\zeta^{-7/10}\;\mu^{3/4}\;\mu_e^{4/5}\;\mu_I^{1/10}\;\ga^{-4/5}\;
\\&&\notag
m^{1/5}\;\dot m^{7/10}\;r^{-3/4}\;s^{7/10}
\;(1.3\times 10^5\;\mrm{g\;cm^{-2}})\\
\gr_c&=&\notag
Z^{-3/10}\;
\zeta^{-11/20}\;\mu^{9/8}\;\mu_e^{7/10}\;\mu_I^{3/20}\;
\ga^{-7/10}\;  \\&&\notag
m^{-7/10}\;\dot{m}^{11/20}\; r^{-15/8}\;s^{11/20}
\;(1.0\times 10^1\;\mrm{g\;cm^{-3}})\\ 
n_{ec}&=&\notag
Z^{-3/10}\;
\zeta^{-11/20}\;\mu^{9/8}\;\mu_e^{-3/10}\;\mu_I^{3/20}\;
\ga^{-7/10}\;  \\&&\notag
m^{-7/10}\;\dot{m}^{11/20}\; r^{-15/8}\;s^{11/20}
\;(6.3\times 10^{24}\;\mrm{cm^{-3}})\\ 
\bar{v}_R&=&\notag
Z^{1/5}\;
\zeta^{-3/10}\;\mu^{-3/4}\;\mu_e^{1/5}\;\mu_I^{-1/10}\;
\ga^{4/5}\;  \\&&\notag
m^{-1/5}\;\dot{m}^{3/10}\; r^{-1/4}\;s^{-7/10}
\;(- 2.0\; \mrm{km\; sec^{-1}})\\ 
\eps_{\gc c}&=&\notag
Z^{4/5}\;
\zeta^{-6/5}\;\mu\;\mu_e^{4/5}\;\mu_I^{-2/5}\;\ga^{-4/5}\;
 \\&&\notag 
m^{-4/5}\;\dot m^{6/5}\;r^{-3}\;s^{6/5}
\;(2.7\times 10^{16}\;\mrm{erg\;cm^{-3}}) \\
P_c&=&\notag
Z^{-1/10}\;
\zeta^{-17/20}\;\mu^{3/8}\;\mu_e^{9/10}\;\mu_I^{1/20}\;
\ga^{-9/10}\;  \\&&\notag
m^{-9/10}\;\dot m^{17/20}\;
r^{-21/8}\;s^{17/20}
\;(3.8\times 10^{16}\;\mrm{erg\;cm^{-3}})\\
T_c&=&\notag
Z^{1/5}\;
\zeta^{-3/10}\;\mu^{1/4}\;\mu_e^{1/5}\;\mu_I^{-1/10}
\;\ga^{-1/5}\;  \\&&\notag
m^{-1/5}\;\dot m^{3/10}\;
r^{-3/4}\;s^{3/10}\;(4.3\times 10^7\;\mrm{K})\\
\tau_c&=&\notag
Z^{3/10}\;\zeta^{-9/20}\;
\mu^{7/8}\;\mu_e^{-1/5}\;\mu_I^{-3/20}\;  
\ga^{-4/5}\; \\&&\notag
m^{1/5}\;\dot{m}^{9/20}\;r^{-3/8}\; 
s^{9/20}\;(1.5\times 10^3).
\ee
%}
The disk zones are labelled by A, B, and C as in Shakura and Sunyaev
(1973). Numerical values based on the above results for different
types of pure, fully ionized plasmas are given in Table \ref{table_1}. For example, this table indicates that, compared to a
pure H disk with same parameter values $(\ga,M,\dot{M})$, the disk becomes
typically thinner by a factor of two, if dominated by heavier
elements.  As one may check by virtue of Saha' s equation
\citep{LaLi03}, considering fully ionized  He, C, Mg, etc. is a good
approximation in the case of ultracompact binaries possessing hot accretion
disks with small radial diameters.  
\par
In addition to the quantities from Eqs. \eqref{e:zone_A} --
\eqref{e:zone_C}, we listed in Table \ref{table_1} the relative
changes of the characteristic time scale for Comptonization \citep{PoSoSu83}
\be
t_\mrm{co}= 
\f{3}{8} \f{m_e c}{\gs_\mrm{T}\,\eps_{\gc c}}.
\ee
Another important time scale is given by the equipartition time for
the energy exchange between fast electrons and slow ions \citep{Sp62} 
\be
t_\mrm{eq}
&=& 
 \f{A}{Z^2\ln \Lambda} 
\left(\f{\mrm{cm}^{-3}}{n_{Ic}}\right)\;
\left(\f{kT_c}{m_e c^2}\right)^{3/2}
(\pot{1.1}{17}\;\mrm{sec})\\
&=& \notag
 \f{A\mu_I}{Z^2\ln \Lambda} 
\left(\f{\mrm{g\;cm}^{-3}}{\gr_{c}}\right)\;
\left(\f{kT_c}{m_e c^2}\right)^{3/2}
(\pot{1.9}{-7}\;\mrm{sec}),
\ee
where $n_{Ic}= \gr_c/(\mu_I m_p)$ is the number density of ions in the 
disk  midplane.  The quantities $t_\mrm{eq}$ and  $t_\mrm{co}$ can 
be compared with, e.g., the time scale of radial motions $t_R=R/\bar{v}_R$, or,
alternatively, with the revolution time scale $t_\Go=2\pi R/\Go$. It
is worthwhile to note that, in contrast to $t_\mrm{eq/co/R}$, the
revolution time scale $t_\Go$ is independent from the chemical
composition. A two-temperature flow regime can develop in the hot, inner
disk region, if the latter becomes optically thin and if
$t_\mrm{eq}\gg t_\Go$ holds \citep{ShLiEa76}.

%%%%%%%%%%%%%%%%%%%%%%%%%%%%%%%%%%%%%%%%%%%%%%%%%%
\section{Boundary (spreading) layer of neutron stars}
\label{s:neutron_stars}
%%%%%%%%%%%%%%%%%%%%%%%%%%%%%%%%%%%%%%%%%%%%%%%%%%

The main difference between accretion
onto black holes and neutron stars is due to the fact that on the
neutron star's surface the kinetic energy of the accreting matter is 
transformed into radiation, whereas in the former case this
energy is absorbed by the black hole. There exist several different
models for  the boundary layers of neutron stars (Sunyaev and Shakura,
1986; Popham and Sunyaev, 2001). In the spreading layer model of
\citet{InSu99}, two bright belts, located equidistant from the equator on the surface of the (slowly rotating) star, emit
radiation corresponding to the local Eddington flux  
\be\label{e:condition}
q_\mrm{E}(\gt)
&=&\f{L_\mrm{E}}{4\pi R_*^2}
\left[1-\left(\f{v_\phi(\gt)}{v_\mrm{K}}\right)^2\right]=
\mu_e\,\f{G M}{R_*^2}\f{m_p c}{\gs_\mrm{T}}
\left[1-\left(\f{v_\phi(\gt)}{v_\mrm{K}}\right)^2\right],
\ee
where $R_*$ denotes the radius of the neutron star, $\gt$ the latitude angle
($\gt=0$ corresponds to the \lq equatorial\rq\space disk mid-plane), 
$v_\mrm{K}=(GM/R_*)^{1/2}$ is Keplerian velocity and $v_{\phi}$ is the rotation velocity on the surface of
the star (to be obtained by solving the boundary layer equations of
Inogamov and Sunyaev, 1999).   
The local Eddington flux $q_\mrm{E}(\gt)$ is determined by the difference of
the gravitational force and the
centrifugal force due to the rotation of the accreting matter on the
surface of star. Hence,
because of $L_\mrm{E}=\mu_e L_\mrm{E,H}$, formula \eqref{e:condition}
implies that the flux that may be emitted from the surface of a
neutron star increases with $\mu_e$ (small deviations from this simple
proportionality may occur due to a weak dependence of
$v_{\phi}$ on the chemical parameters $\mu_e$ and
$\mu$). Therefore, in the case of He or metal rich accretion flows,
this flux is increased by a factor of two (at same values of $\dot
M$), compared to a pure H disk. Moreover, similar to the scale
height $H$ of the disk, the meridional height  $H_\mrm{SL}$ of the
boundary (spreading) layer, which can be estimated from the  
energy balance \citep{InSu99}
\be\label{e:LSL}
\f{\dot M v_\mrm{K}^2}{2}=L_\mrm{SL}
=L_\mrm{E}\f{H_\mrm{SL}}{R_*}
=\mu_e L_\mrm{E,H}\f{H_\mrm{SL}}{R_*},
\ee
decreases at same values of $\dot
M$, if the disk is dominated by elements heavier than H. 

\paragraph{Acknowledgments}
The authors are grateful to N. Shakura for his careful reading
of the manuscript and several helpful remarks.

%%%%%%%%%%%%%%%%%%%%%%%%%%%%%%%%%%%%%%%%%%%%%%%%%%

\appendix
%%%%%%%%%%%%%%%%%%%%%%%%%%%%%%%%%%%%%%%%%%%%%%%%%%
\section{Paszy\'nsky-Wiita potential} 
\label{a:relativistic_effects}
%%%%%%%%%%%%%%%%%%%%%%%%%%%%%%%%%%%%%%%%%%%%%%%%%%

As discussed by \citet{PaWi80}, the effects of
general relativity can be modeled by replacing the Newtonian
potential $\Phi=-{G M}/r$  with the modified potential  
\be\label{e:modified-potential} 
\hat\Phi(R,z)=-\f{G M}{r-r_G},
\ee 
corresponding to an efficiency factor $\zeta=1/16$, in contrast to the
Newtonian value $\zeta=1/12$. 
In order to simplify subsequent formulae, it is convenient to define three dimensionless correction factors by
\bse
\begin{alignat}{2}
\chi_0&\equiv \f{R}{R-r_G}&
&=\f{r}{r-1/3},\\
\chi_1&\equiv\f{3}{2}-\f{R+\sqrt{3R r_G}}{2(R-r_G)}&
&=\f{3}{2}-\f{r+\sqrt{r}}{2(r-1/3)},\\
\chi_2&\equiv\f{R(3R-r_G)}{6(R-r_G)^3}\left(2R-\sqrt{3R
r_G}-3r_G\right)&
&=\f{r(3r-1/3)}{6(r-1/3)^3}\left(2r-\sqrt{r}-1\right).
\end{alignat}
\ese 
As can be seen from these factors, relativistic corrections are
important only in the innermost parts of the disk. Therefore, we will
confine ourselves here to giving the results for zones A and B.

%%%%%%%%%%%%%%%%%%%%%%%%%%%%%%%%%%%%%%%%%%%%%%%%%% 
\paragraph{Zone A:}  
\label{a-s:results_zone_A} 
%%%%%%%%%%%%%%%%%%%%%%%%%%%%%%%%%%%%%%%%%%%%%%%%%%

In the case of dominating radiation pressure, $P_g\ll P_\gc$, and
dominating Thomson scattering, $\gk_\mrm{T} \gg\gk_\mrm{ff}$, 
one finds
\be\label{e:PW-results_A}
\hat H(R)&=&
\chi_0^{-2}\;\chi_2\;H\\
\hat \Gs(R)&=&\notag
\chi_0^{2}\;\chi_1\;\chi_2^{-2}\;\Gs\\
\hat \gr_c(R)
&=&\notag
\chi_0^{4}\;\chi_1\;\chi_2^{-3}\; \gr_c\\
\hat n_{e c}(R)&=&\notag
\chi_0^{4}\;\chi_1\;\chi_2^{-3}\; n_{e c}\\
\hat{\bar v}_R(R)&=&\notag
\chi_0^{-2}\;\chi_1^{-1}\;\chi_2^{2}\;{\bar v}_R\\
\hat \eps_{\gc c}(R)&=&\notag
\chi_0^{2}\;\chi_1\;\chi_2^{-1}\; \eps_{\gc c}\\
\hat P_c(R)&=&\notag
\chi_0^{2}\;\chi_1\;\chi_2^{-1}\;P_c\\
\hat T_c(R)&=&\notag
\chi_0^{1/2}\;\chi_1^{1/4}\;\chi_2^{-1/4}\;T_c\\
\hat\tau_c(R)&=&\notag
\chi_0^{25/8}\;\chi_1^{17/16}\;\chi_2^{-49/16}\;\tau_c,
\ee
where on the rhs. the results from
Eqs. \eqref{e:zone_A} are to be inserted (here and
below non-hatted quantities refer to the expressions derived for the
case of the Newtonian potential). 
 
%%%%%%%%%%%%%%%%%%%%%%%%%%%%%%%%%%%%%%%%%%%%%%%%%%
\paragraph{Zone B:}
\label{a-s:results_zone_B}
%%%%%%%%%%%%%%%%%%%%%%%%%%%%%%%%%%%%%%%%%%%%%%%%%%

In the case of dominating gas pressure, $P_g\gg P_\gc$, and
dominating Thomson scattering, $\gk_\mrm{T} \gg\gk_\mrm{ff}$, 
one finds
\be\label{e:PW-results_B}
\hat H(R)&=&
\chi_0^{-1}\;\chi_1^{1/10}\;\chi_2^{1/10}\;H\\
\hat \Gs(R)&=&\notag
\chi_1^{4/5}\;\chi_2^{-1/5}\;\Gs\\
\hat \gr_c(R)
&=&\notag
\chi_0\;\chi_1^{7/10}\;\chi_2^{-3/10}\; \gr_c\\
\hat {n}_{ec}(R)&=&\notag
\chi_0\;\chi_1^{7/10}\;\chi_2^{-3/10}\;{n_e}_c\\
\hat{\bar v}_R(R)&=&\notag
\chi_1^{-4/5}\;\chi_2^{1/5}\;{\bar v}_R\\
\hat \eps_{\gc c}(R)&=&\notag
\chi_1^{4/5}\;\chi_2^{4/5}\;\eps_{\gc c}\\
\hat P_c(R)&=&\notag
\chi_0\;\chi_1^{9/10}\;\chi_2^{-1/10}\;P_c\\
\hat T_c(R)&=&\notag
\chi_1^{1/5}\;\chi_2^{1/5}\;T_c\\
\hat\tau_c(R)&=&\notag
\chi_0^{1/2}\;\chi_1^{4/5}\;\chi_2^{-7/10}\;\tau_c,
\ee
where on the rhs. one has to insert the results from
Eqs. \eqref{e:zone_B}. 
\par
A dimensionless transition radius $r_{AB}$, separating zones $A$
and $B$, can be determined from the condition $\hat T^A(r_{AB})=\hat T^B(r_{AB})$; i.e., $r_{AB}$ is given by
the solution of the equation 
\be\label{a-e:PW-r-A-B-condition}
r ^{21/16} s^{-1}
\chi_0^{5/4}\;\chi_1^{1/8}\;\chi_2^{-9/8}
=
47.3\;
\zeta^{-1} \;\mu^{1/2}\;\mu_e^{-1/8}\;\ga^{1/8}\; m^{1/8}\;\dot m.
\ee 
The function on the lhs. attains its {\em minimum} value $14.8$ at
$r=2.54$. Consequently, zone A exists only if  
\be\label{a-e:radius-A-B}
0.02< \mu^{1/2}\;\mu_e^{-1/8}\;\ga^{1/8}\; m^{1/8}\;\dot m,
\ee
where we have already inserted the efficiency factor $\zeta=1/16$. 

\end{document}